\documentclass[conference]{IEEEtran}
\IEEEoverridecommandlockouts
\usepackage{cite}
\usepackage{amsmath,amssymb,amsfonts}
\usepackage{algorithmic}
\usepackage{graphicx}
\usepackage{textcomp}
\usepackage{mathrsfs}
\usepackage{subfigure}
\usepackage{bm}
\usepackage{xcolor}
\def\BibTeX{{\rm B\kern-.05em{\sc i\kern-.025em b}\kern-.08em
    T\kern-.1667em\lower.7ex\hbox{E}\kern-.125emX}}
\begin{document}

\title{Digital Twin-based Cooperative Autonomous Driving in Smart Intersections: A Multi-Agent Reinforcement Learning Approach}

\author{
    \IEEEauthorblockN{
        Taoyuan Yu\IEEEauthorrefmark{1},
        Kui Wang\IEEEauthorrefmark{1},
        Zongdian Li\IEEEauthorrefmark{1},
        Tao Yu\IEEEauthorrefmark{1},
        Kei Sakaguchi\IEEEauthorrefmark{1},
        and Walid Saad\IEEEauthorrefmark{2}
    }
    \IEEEauthorblockA{
        \IEEEauthorrefmark{1}Department of Electrical and Electronic Engineering, Institute of Science Tokyo, Tokyo, TYO 152-8550, Japan \\
        \IEEEauthorrefmark{2}Bradley Department of Electrical and Computer Engineering, Virginia Tech, Arlington, VA 22203, USA \\
        Email: \{yuty, kuiw, lizd, yutao, sakaguchi\}@mobile.ee.titech.ac.jp, walids@vt.edu
    }
}

\maketitle

\begin{abstract}
Unsignalized intersections pose safety and efficiency challenges due to complex traffic flows and blind spots. In this paper, a digital twin (DT)-based cooperative driving system with roadside unit (RSU)-centric architecture is proposed for enhancing safety and efficiency at unsignalized intersections. The system leverages comprehensive bird-eye-view (BEV) perception to eliminate blind spots and employs a hybrid reinforcement learning (RL) framework combining offline pre-training with online fine-tuning. Specifically, driving policies are initially trained using conservative Q-learning (CQL) with behavior cloning (BC) on real datasets, then fine-tuned using multi-agent proximal policy optimization (MAPPO) with self-attention mechanisms to handle dynamic multi-agent coordination.  The RSU implements real-time commands via vehicle-to-infrastructure (V2I) communications. Experimental results show that the proposed method yields failure rates below 0.03\% coordinating up to three connected autonomous vehicles (CAVs), significantly outperforming traditional methods. In addition, the system exhibits sub-linear computational scaling with inference times under 40 ms. Furthermore, it demonstrates robust generalization across diverse unsignalized intersection scenarios, indicating its practicality and readiness for real-world deployment.
\end{abstract}


\section{Introduction}
Intersection management remains a critical bottleneck in intelligent transportation systems (ITS) due to intersection complexity and uncertainty \cite{b1}. According to the Federal Highway Administration (FHWA) and National Highway Traffic Safety Administration (NHTSA), intersection-related fatalities constitute a significant portion of traffic accident deaths, with unsignalized intersections accounting for 68\% in 2024 \cite{b2,b3}. Blind spots and unclear interaction protocols make unsignalized intersections particularly dangerous. To address these challenges, the concept of a digital twin (DT) provides a promising solution by creating real-time virtual replicas of physical intersections, providing global perception and intelligent coordination beyond the limited sensing capabilities of individual vehicles \cite{b4,b5}.

Mixed-traffic scenarios involving autonomous vehicles (AVs) and human-driven vehicles (HDVs) are becoming increasingly common thereby increasing the complexity of coordination among traffic participants.  Vehicle-to-everything (V2X) communications technologies, including vehicle-to-vehicle (V2V), vehicle-to-infrastructure (V2I), vehicle-to-pedestrian (V2P), and vehicle-to-network (V2N), can help enhance traffic safety and efficiency \cite{b6, b7}. Among these, V2I communications play a central role in DT systems as they allow real-time synchronization between physical vehicles and roadside units (RSUs), thereby supporting cooperative driving strategies and transforming traditional intersection infrastructure into intelligent control centers \cite{b8}.

\begin{figure*}[t]
     \centerline{\includegraphics[width=0.9 \textwidth]{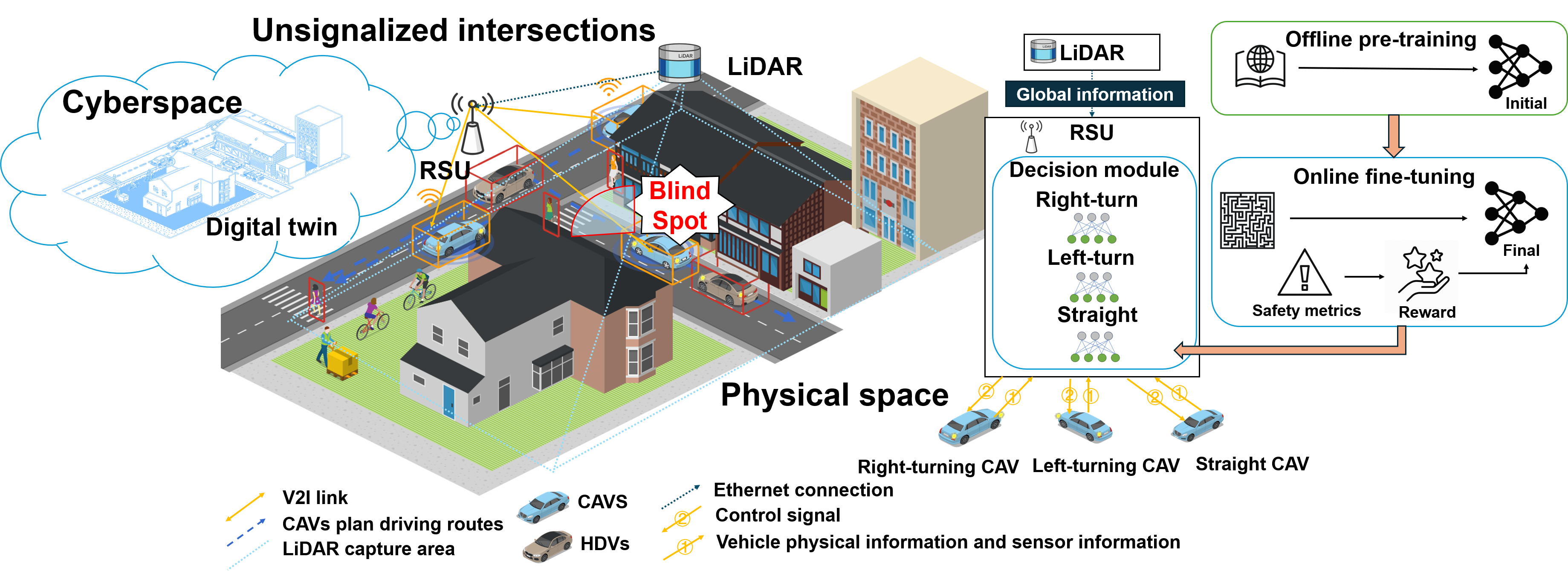}}
     \caption{High-level architecture of the DT cooperative system}
     \label{fig: sysd}
\end{figure*}

Leveraging V2I communications, DT systems have been applied to intersection management with various architectures. For example, in \cite{b9} and \cite{b10}, the authors developed RSU-based DTs for continuous traffic monitoring and real-time analysis through cloud computing. However, these methods focus on general traffic monitoring and basic perception enhancement, without addressing blind spots and occlusions at intersections. The work in \cite{b11} addresses intersection occlusions but the solution of \cite{b11} is limited by local vehicle sensing and cannot achieve complete blind spot elimination. Despite the potential of DT technology to provide complete environmental awareness, current implementations \cite{b9, b10, b11} neither eliminate blind spots through global bird-eye-view (BEV) nor support cooperative driving strategies in occluded areas.

To leverage comprehensive DT perception, various intersection coordination algorithms have been explored in \cite{b12, b13, b14, b15, b16}. Traditional methods use optimization and game-theoretic algorithms to manage traffic flow \cite{b12, b13}, but lack adaptability in dynamic environments. Multi-agent reinforcement learning (MARL) has been proposed as an effective solution for flexible and scalable coordination under partial observability \cite{b14, b15, b16}. Recent works further improve MARL by incorporating self-attention mechanisms to enhance inter-agent communication and decision-making. However, most MARL models adopt uniform policies across agents and fail to model diverse driving intents such as left-turn, straight, or right-turn maneuvers. In addition, these models are rarely evaluated under various vehicle densities, making their robustness in dynamic traffic conditions uncertain. Critically, current MARL methods fail to exploit DT’s global perception for blind-spot elimination, leaving a gap in ITS research.

The main contribution of this paper is to address the above limitations by developing a novel DT-based cooperative driving system for unsignalized intersections. The system leverages RSU-mounted LiDAR to construct comprehensive BEV perception to eliminate blind spots, creating a real-time digital replica of the intersection environment. Our method leverages a centralized MARL decision module with role-specific policy networks and self-attention mechanisms. This method allows robust cooperative driving for various numbers of vehicles. Through a hybrid learning framework combining offline pre-training with online fine-tuning, the system develops decision-making capabilities that can be effectively deployed in real-world scenarios. This design achieves significant improvements in blind spot elimination, system adaptability, and traffic efficiency. In summary, our key contributions include:

\begin{itemize}
    \item We develop a DT-based MARL framework eliminating blind spots via RSU global perception at unsignalized intersections.
    \item We introduce role-specific policy networks with self-attention mechanisms to enable adaptive coordination among connected autonomous vehicles (CAVs).
    \item We propose a hybrid offline-online reinforcement learning method to ensure robust and efficient policy learning.
    \item We conduct extensive experiments demonstrating system effectiveness and generalization across diverse scenarios.
\end{itemize}

The rest of this paper is organized as follows: Section II presents the DT system architecture. Section III details the proposed algorithm. Section IV discusses experimental results. Section V concludes the paper and outlines future work.

\begin{figure*}[t]
     \centerline{\includegraphics[width=1.0 \textwidth]{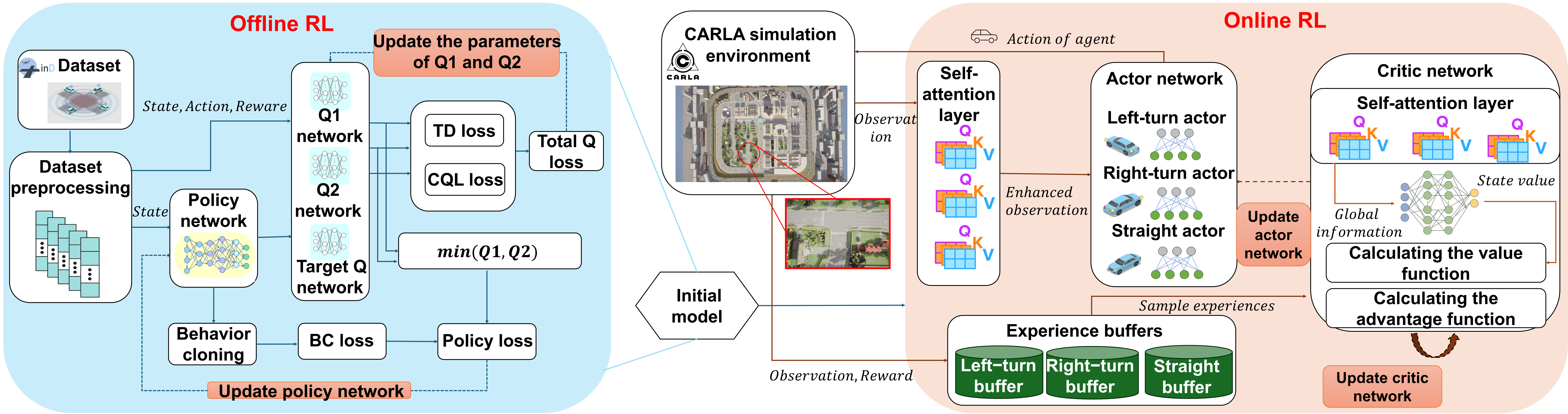}}
     \caption{Offline-online hybrid RL algorithm framework design}
     \label{fig:af}
\end{figure*}

\section{RSU-CAVs Cooperative System}

We consider the DT-based cooperative driving system architecture is shown in Fig.~\ref{fig: sysd}. The system establishes real-time synchronization between the physical intersection and its DT in cyberspace. RSU-mounted LiDAR sensors provide comprehensive BEV perception which, in turn, can help eliminate blind spots for global traffic monitoring. In contrast to traditional vehicle-centric methods focused on individual vehicles, this DT-based system facilitates centralized decision-making for multiple CAVs at unsignalized intersections. The RSU's global perception overcomes individual vehicle sensor limitations, enabling a cooperative driving strategy designed to minimize potential conflicts and maximize intersection throughput.

To effectively manage the intersection's complexities and uncertainties, the RSU employs decision-making policies developed through a two-stage learning method. Given the dynamic and partially observable nature of traffic environments, RL provides a framework for sequential decision-making under uncertainty, modeled as a partially observable Markov decision process (POMDP). The training process begins with offline RL on real-world traffic dataset to establish foundational driving strategies, followed by online RL in simulated environments to enhance adaptability and robustness. This hybrid paradigm ensures that the resulting policies can handle diverse traffic scenarios while maintaining safety constraints. Once deployed, the RSU leverages these trained policies within its DT system to make real-time decisions, minimizing onboard compute requirements for CAVs while ensuring low-latency response \cite{b17}.

To address the challenges of blind spots, and limited onboard sensing at unsignalized intersections, we introduce a DT-based cooperative system. As CAVs approach the intersection, they are simultaneously represented in the DT through V2I communications. The DT maintains real-time synchronization between physical vehicles and their digital counterparts, enabling the RSU to make decisions based on complete traffic state information. Using real-time data, the RSU determines each vehicle's driving role within the DT. Subsequently, the RSU leverages pre-loaded role-based strategy networks in the centralized decision module to compute control signals. These signals are transmitted in real-time to the corresponding CAVs through V2I communications. Concurrently, the DT continuously monitors traffic conditions, including the states and predicted movements of all traffic participants, traffic flow smoothness, and abnormal situations. This synchronization between physical space and cyberspace provides necessary real-time inputs for the decision networks and facilitates performance evaluation.

\section{Hybrid Reinforcement Learning Framework}

As shown in Fig.~\ref{fig:af}, we propose a DT-based two-stage learning framework to develop cooperative driving strategies for unsignalized intersections. This method first employs offline pre-training on collected datasets, using offline RL to acquire foundational driving skills and traffic priors. Subsequently, online fine-tuning within the CARLA simulator \cite{b18} allows agents to adapt to dynamic environments. This hybrid method combines the safety of offline RL with the adaptability of online RL, ensuring that the trained model can achieve real-time decision-making within the RSU's DT system. This section details the methods for offline pre-training and online fine-tuning.

\subsection{Observation Space}

At each time step $t$, the state space $\bm{s}(t)$ encompasses all traffic participants monitored by the RSU. The RSU uses global information to construct individual observation vectors $\bm{o}(t)$ for each CAV, capturing partially observable and potentially noisy representations:
\begin{equation}
    \bm{o}(t) = [\bm{o}_{\text{core}}, \bm{o}_{\text{veh}}, \bm{o}_{\text{ped}}, \bm{o}_{\text{role}}, \bm{o}_{\text{ctx}}],
\end{equation}
where $\bm{o}_{\text{core}}$ includes ego-vehicle speed, global position, heading angle, and junction occupancy; $\bm{o}_{\text{veh}}$ includes relative positions and velocities of nearby vehicles; $\bm{o}_{\text{ped}}$ represents pedestrian detection, distance, and angle; $\bm{o}_{\text{role}}$ encodes the agent’s driving role; and $\bm{o}_{\text{ctx}}$ contains scenario identifiers.

\subsection{Action Space}

We define a unified two-dimensional continuous action space $\mathcal{A}$ for the vehicle. It is structured as:
\begin{equation}
\bm{a}(t) = [\bm{a}_{\text{acc}}, \bm{a}_{\text{steer}}] \in \mathbb{R}^2
\end{equation}
where $\bm{a}_{\text{acc}}$ is longitudinal acceleration and $\bm{a}_{\text{steer}}$ is steering angular velocity. During offline pre-training, actions are estimated from consecutive state transitions, as ground-truth controls are unavailable. During online fine-tuning, actions are directly predicted by the policy network.

\subsection{Reward Function}

To enable cooperative driving, we design a structured reward function $\mathcal{R}_{\text{online}}(\bm{s}(t), \bm{a}(t), \bm{s}(t+1))$ to translate high-level objectives into real-time feedback. The overall reward $r(t)$ is defined as:
\begin{equation}
    r(t) = \sum w_k r_k(\bm{s}(t), \bm{a}(t), \bm{s}(t+1)),
\end{equation}
where $r_i$ represents individual reward components and $w_i$ captures the corresponding weight. The reward terms include:
\begin{equation}
\begin{aligned}
r_i \in \{ 
    & r_{\text{safety}},\ r_{\text{eff}},\ r_{\text{comfort}}, \\
    & r_{\text{task}},\ r_{\text{yield}},\ r_{\text{coop}},\ r_{\text{penalty}} 
\}
\end{aligned}
\end{equation}
where $r_{\text{safety}}$ penalizes hazardous behaviors based on metrics like minimum time-to-collision (TTC); $r_{\text{eff}}$ encourages speed compatible with traffic flow; $r_{\text{comfort}}$ penalizes large acceleration changes; $r_{\text{task}}$ rewards agents reaching navigation targets cooperatively; $r_{\text{yield}}$ and $r_{\text{coop}}$ reward compliance with traffic rules and cooperation; and $r_{\text{penalty}}$ severely penalizes collisions or timeouts. Each term is scaled by its corresponding weight $w_k$, where $w_{\text{safety}}$ and $w_{\text{penalty}}$ are typically assigned larger values due to their critical importance.

\subsection{Offline Pre-training: Networks and Algorithm}

The primary goal of offline pre-training is to provide high-quality initialization for online fine-tuning. The model is trained independently for each driving role using subsets of the InD dataset \cite{b19}, partitioned based on vehicle intentions.

For each subset, we employ an offline RL algorithm combining conservative Q-learning (CQL) and behavior cloning (BC) \cite{b20,b21} in an actor-critic framework. The critic uses twin Q-networks $Q_{\theta_{i,1}}$, $Q_{\theta_{i,2}}$ with target networks to stabilize learning and reduce overestimation, optimized as:
\begin{equation}
\begin{aligned}
    L_Q(\theta_{i,j}) = & \mathbb{E}_{(\bm{o},\bm{a},r,\bm{o}') \sim \mathcal{D}_{\text{role}=i}} \left[ \frac{1}{2}(Q_{\theta_{i,j}}(\bm{o},\bm{a}) - y)^2 \right] \\
    & + \alpha_{\text{CQL}} L_{\text{CQL\_reg}}(\theta_{i,j})
\end{aligned}
\end{equation}
Here, $y = r + \gamma (1 - d) \min_j Q_{\theta_{i,j}'}(\bm{o}', \pi_{\phi_i}(\bm{o}'))$ is the temporal difference (TD) target.

The policy network $\pi_{\phi_i}$ minimizes BC loss and maximizes conservative Q-values:
\begin{equation}
\begin{aligned}
    L_{\pi}(\phi_i) = & \mathbb{E}_{\bm{o} \sim \mathcal{D}_{\text{role}=i}} \left[ - \min_{j=1,2} Q_{\theta_{i,j}}(\bm{o}, \pi_{\phi_i}(\bm{o})) \right] \\
    & + \lambda_{\text{BC}} \mathbb{E}_{(\bm{o},\bm{a}) \sim \mathcal{D}_{\text{role}=i}} \left[ \lVert \pi_{\phi_i}(\bm{o}) - \bm{a} \rVert^2 \right]
\end{aligned}
\end{equation}
where $\alpha_{\text{CQL}}$ and $\lambda_{\text{BC}}$ denote hyperparameters controlling the strength of CQL regularization and BC imitation.

Role-specific actor $\pi_{\phi_{\text{role}}}$ and critic $Q_{\theta_{\text{role}}}$ networks are implemented as multi-layer perceptrons (MLPs). Self-attention is omitted at this stage to ensure robust training stability. The resulting pre-trained weight are reused during online fine-tuning to improve performance and accelerate adaptation.

\begin{figure}[t]
\centering
\subfigure[]{
\begin{minipage}[b]{0.4\textwidth}
\includegraphics[width=1\textwidth]{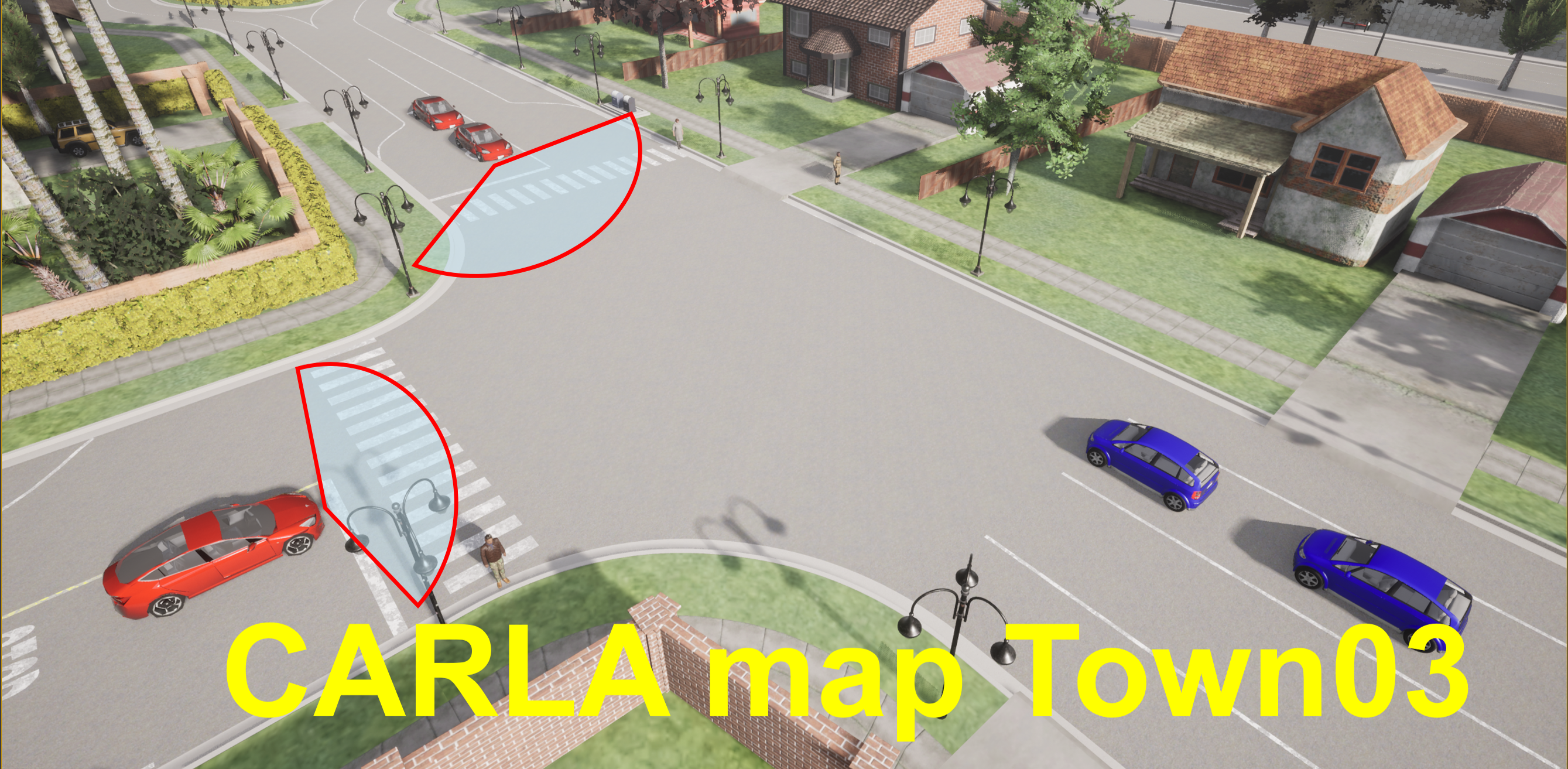} 
\end{minipage}
}
\subfigure[]{
\begin{minipage}[b]{0.4\textwidth}
\includegraphics[width=1\textwidth]{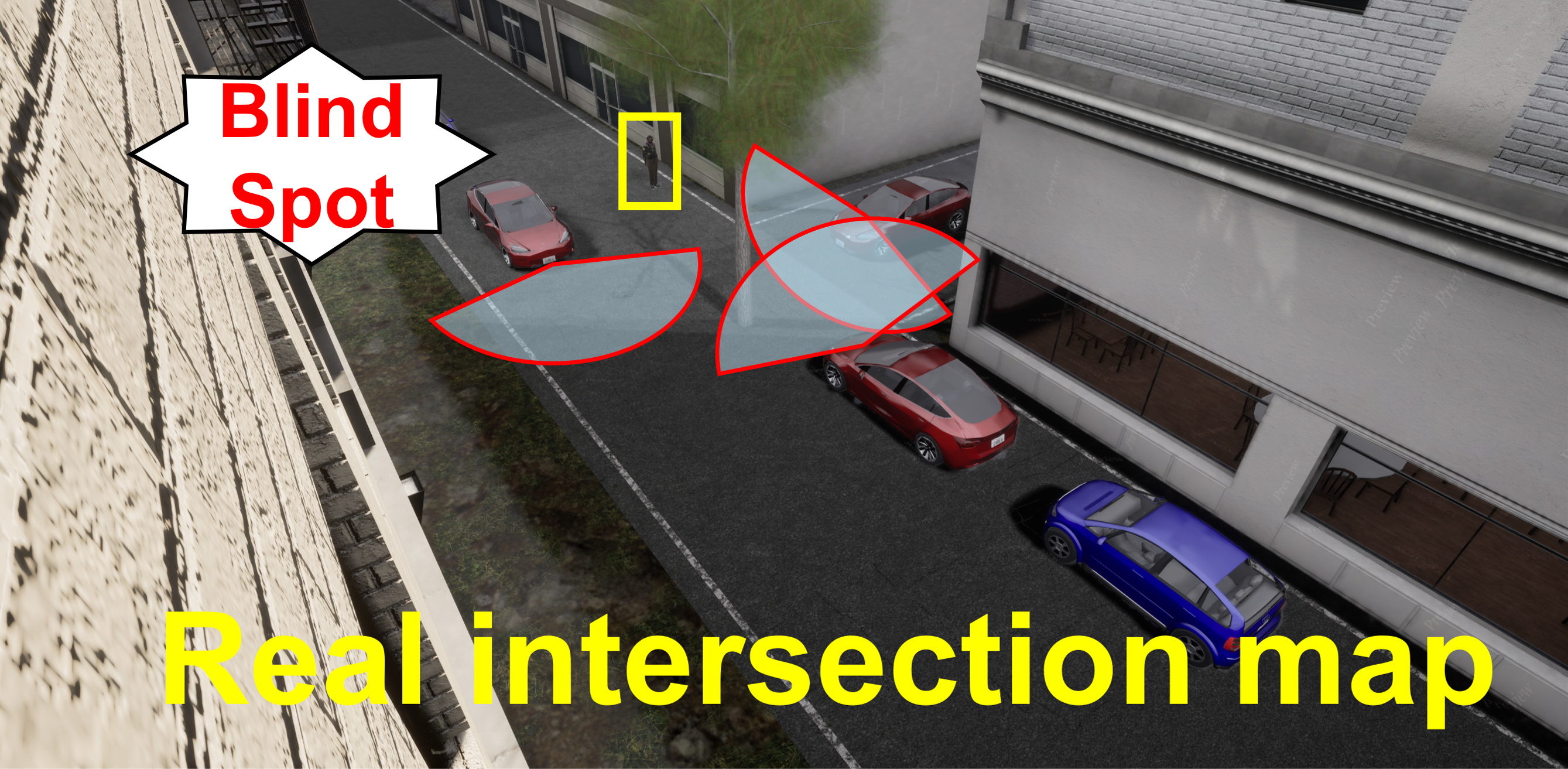} 
\end{minipage}
}
\caption{Experimental scenario and generalization scenario settings (a) CARLA example map, (b) Real intersection map}
\label{fig:tu}
\end{figure}

\subsection{Online Fine-tuning: Networks and Algorithm}

The online fine-tuning adopts multi-agent proximal policy optimization (MAPPO) \cite{b22}, integrating role-specific networks ($\pi_{\phi_{\text{left}}}, \pi_{\phi_{\text{straight}}}, \pi_{\phi_{\text{right}}}$) with a shared critic network $V_\psi$.

\begin{figure}[t]
     \centerline{\includegraphics[width=0.45 \textwidth]{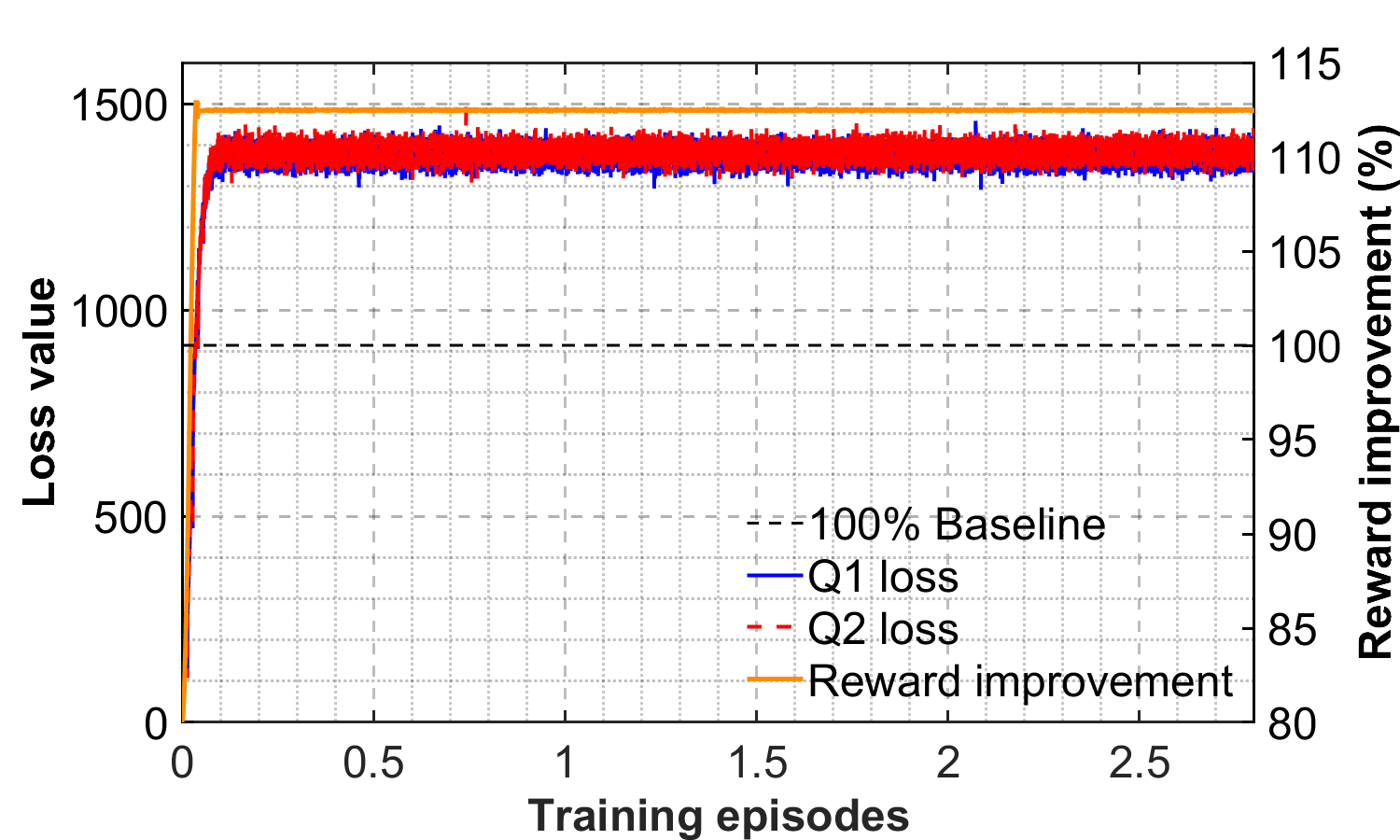}}
     \caption{Offline pre-training results}
     \label{fig:4}
\end{figure}

To capture dynamic interactions, we augment both actor and critic networks with multi-head self-attention (MHSA). MHSA allows the model to jointly attend to information from different representation subspaces at different positions. The scaled dot-product attention is defined as:
\begin{equation}
\label{eq:attention}
\text{A}(Q, K, V) = \text{softmax}\left(\frac{QK^\top}{\sqrt{d_k}}\right) V
\end{equation}
where $Q$, $K$, and $V$ denote the query, key, and value matrices. MHSA computes multiple attention heads in parallel, and concatenates their outputs to form the final embedding, capturing dependencies among observation features.

Online learning proceeds via an interact-learn loop. Agents generate trajectories:
\begin{equation}
\tau = \left\{(\bm{o}_t, \bm{a}_t, r_{t+1}, V_\psi(\bm{o}_t), \log \pi_{\phi_{\text{role}}}(\bm{a}_t \mid \bm{o}_t))\right\}_{t=0}^{T}
\end{equation}

Advantage estimates $\hat{A}_t^{\text{GAE}}$ and returns $\hat{R}_t$ are computed using generalized advantage estimation (GAE), based on temporal-difference (TD) errors $\delta_t$ calculated from critic values as:
\begin{equation}
\delta_t = r_{t+1} + \gamma V_\psi(\bm{o}_{t+1}) - V_\psi(\bm{o}_t)
\end{equation}

Prioritized experience replay (PER) samples transitions based on priorities proportional to absolute TD errors, with importance sampling (IS) weight correcting sampling bias:
\begin{equation}
w_t = \left( \frac{1}{B \cdot P(t)} \right)^\beta
\end{equation}
where $B$ is the replay buffer size, and $\beta$ controls IS correction strength.

Each role-specific actor $\pi_{\phi_{\text{role}}}$ is trained using the following weighted objective, which includes the PPO clipped surrogate loss and an entropy bonus $S[\cdot]$:
\begin{equation}
\begin{aligned}
L^{\text{CLIP+S}}(\phi_{\text{role}}) = \mathbb{E}_{t \sim \text{PER}} \Big[ w_t \big(
& - L_t^{\text{CLIP}}(\phi_{\text{role}}) \\
& - c_2 \cdot S[\pi_{\phi_{\text{role}}}](\bm{o}_t) \big) \Big]
\end{aligned}
\end{equation}

The PPO surrogate loss $L_t^{\text{CLIP}}$ is defined as:
\begin{equation}
L_t^{\text{CLIP}} = \min \left( r_t \hat{A}_t, \text{clip}(r_t, 1 - \epsilon, 1 + \epsilon) \hat{A}_t \right)
\end{equation}
where $\epsilon$ is the PPO clipping hyperparameter, and $r_t$ represents the probability ratio between current and old policies.

\section{Experiments and Analysis}

Experiments were conducted in synchronous mode using the CARLA simulator with Unreal Engine. The main test scenario is an unsignalized intersection in Town03, as shown in Fig.~\ref{fig:tu}. In each episode, our system controls 1 to 3 CAVs (red), while background vehicles (blue) are controlled by CARLA’s Traffic Manager. Pedestrians are added to simulate realistic urban conditions. For generalization evaluation, we deploy the model on a real intersection map based on the Institute of Science Tokyo campus. The RSU maintains a global state via BEV perception and uses the fine-tuned decision model to compute control commands, which are sent to CAVs through simulated V2I communication.

\begin{figure}[t]
\centering
\subfigure[]{
\begin{minipage}[b]{0.4\textwidth}
\includegraphics[width=1.\textwidth]{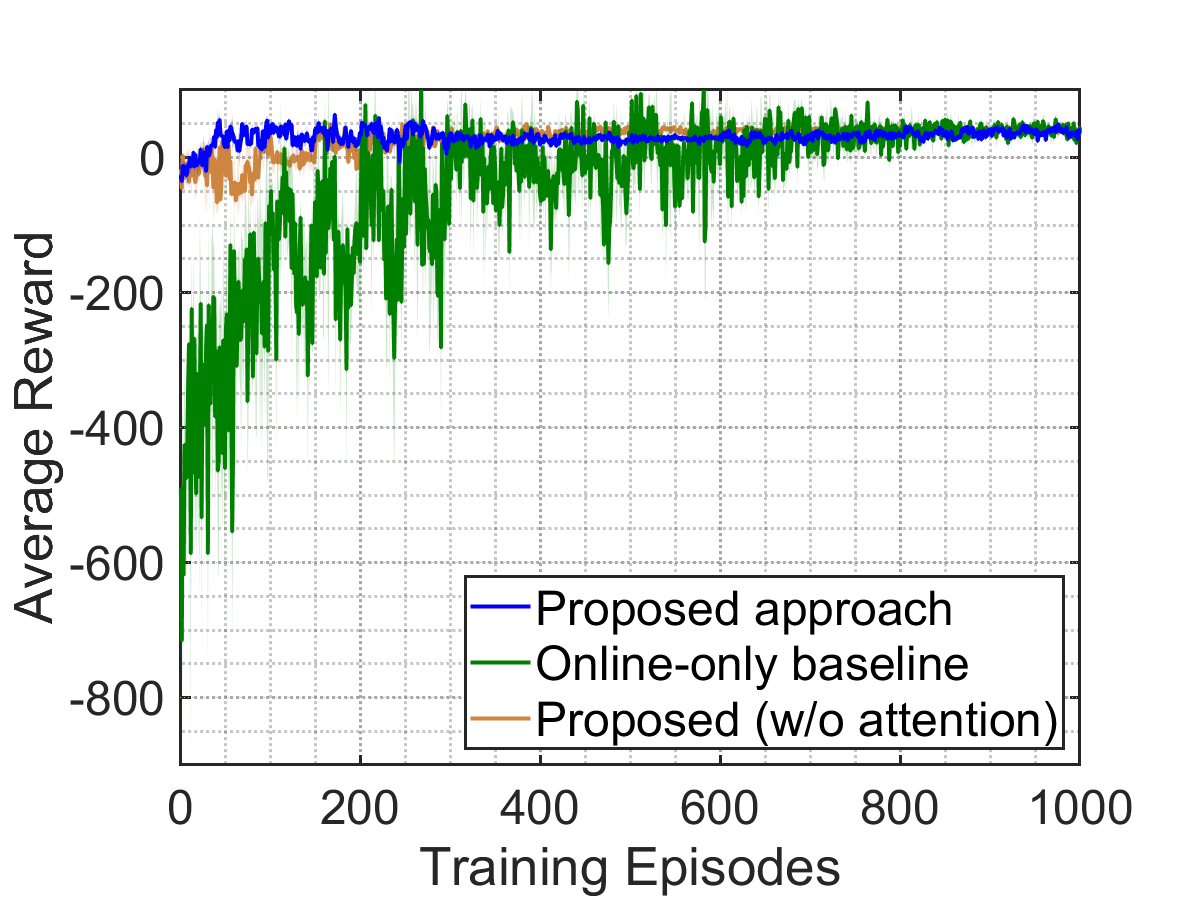} 
\end{minipage}
}
\subfigure[]{
\begin{minipage}[b]{0.4\textwidth}
\includegraphics[width=1\textwidth]{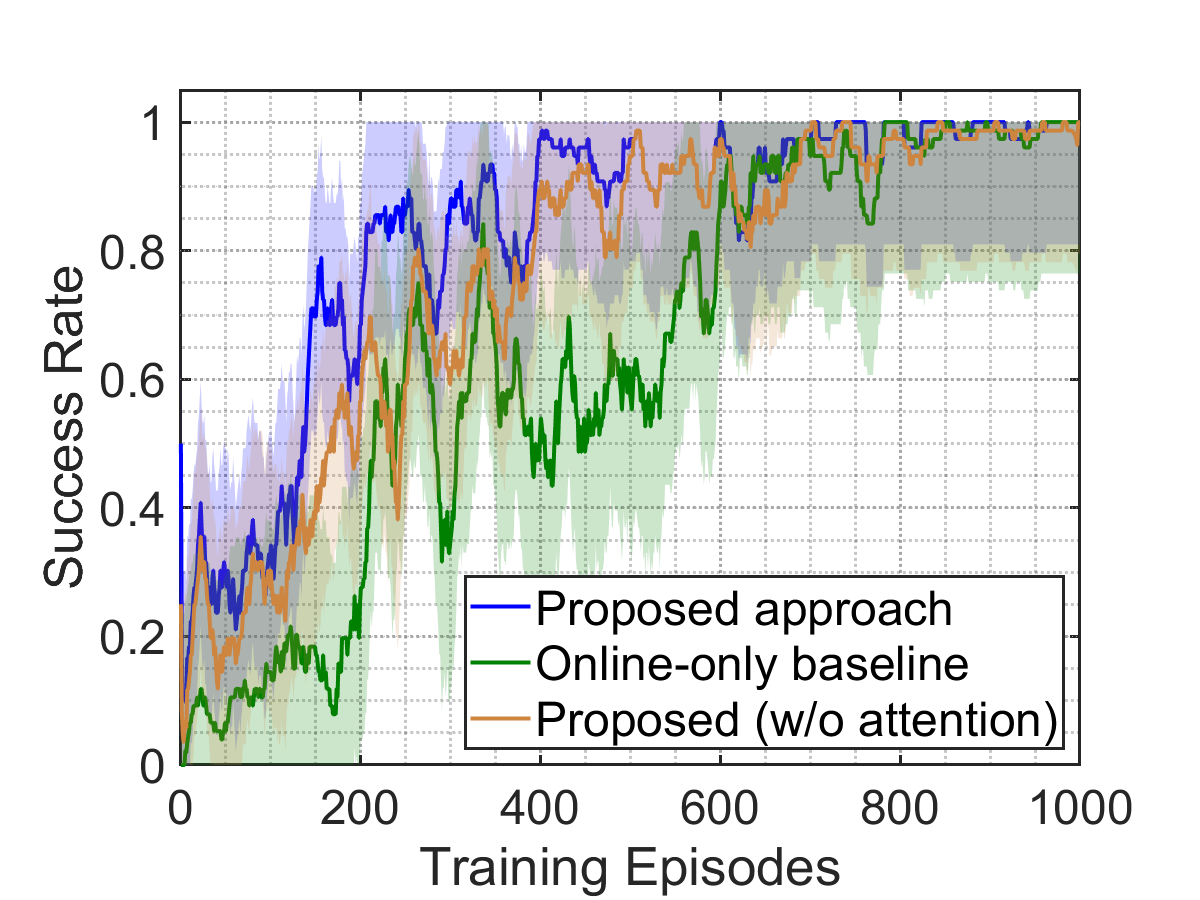} 
\end{minipage}
}
\caption{Comparison of training performance across different approaches, (a) reward, (b) success rate.}
\label{fig:5}
\end{figure}

\subsection{Baselines and Evaluation Metrics}

To assess the contribution of each component, we compare our model with several baselines. First, two ablated variants are considered: (1) an online-only MAPPO baseline trained directly; and (2) a variant with offline pre-training but without self-attention or role-specific policies. Both share the same architecture and hyperparameters as full model, isolating the effects of offline pre-training and self-attention respectively. Second, we include Autoware Universe \cite{b23}, a rule-based autonomous driving stack, configured to control a single vehicle. All approaches are evaluated in terms of convergence speed, failure rate, and average travel time.

\subsection{Offline Pre-training Results}

The offline pre-training phase aims to extract driving priors from the InD dataset to initialize the model for online fine-tuning. Fig.~\ref{fig:4} shows the Q1/Q2 losses and reward improvement over training. The steadily converging losses indicate stable learning of state-action values, while the reward metric stabilizes around 112\%, surpassing the 100\% baseline. This confirms that the policy learned via CQL combined with BC not only imitates but also improves upon average dataset behavior, offering a strong initialization for the online stage.

\begin{table}[t]
\centering
\caption{Performance comparison summary}
\label{tab:performance_comparison_simple}
\scriptsize
\begin{tabular}{|c|c|c|}
\hline
\textbf{approach / Scenario} & \textbf{Failure rate (\%)} & \textbf{Avg. time (s)} \\
\hline
Ours (1 Agent, Town03) & 0.01 & 5.52 \\
Ours (2 Agent, Town03) & 0.03 & 5.49 \\
Ours (3 Agent, Town03) & 0.02 & 5.25 \\
\hline
Autoware (1 Agent, Town03) & 5.31 & 5.77 \\
\hline
Ours (3 Agent, Real Map) & 0.02 & 5.15 \\
\hline
\end{tabular}
\end{table}

\begin{figure}[t]
\centering
\subfigure[]{
\begin{minipage}[b]{0.45\textwidth}
\includegraphics[width=1\textwidth]{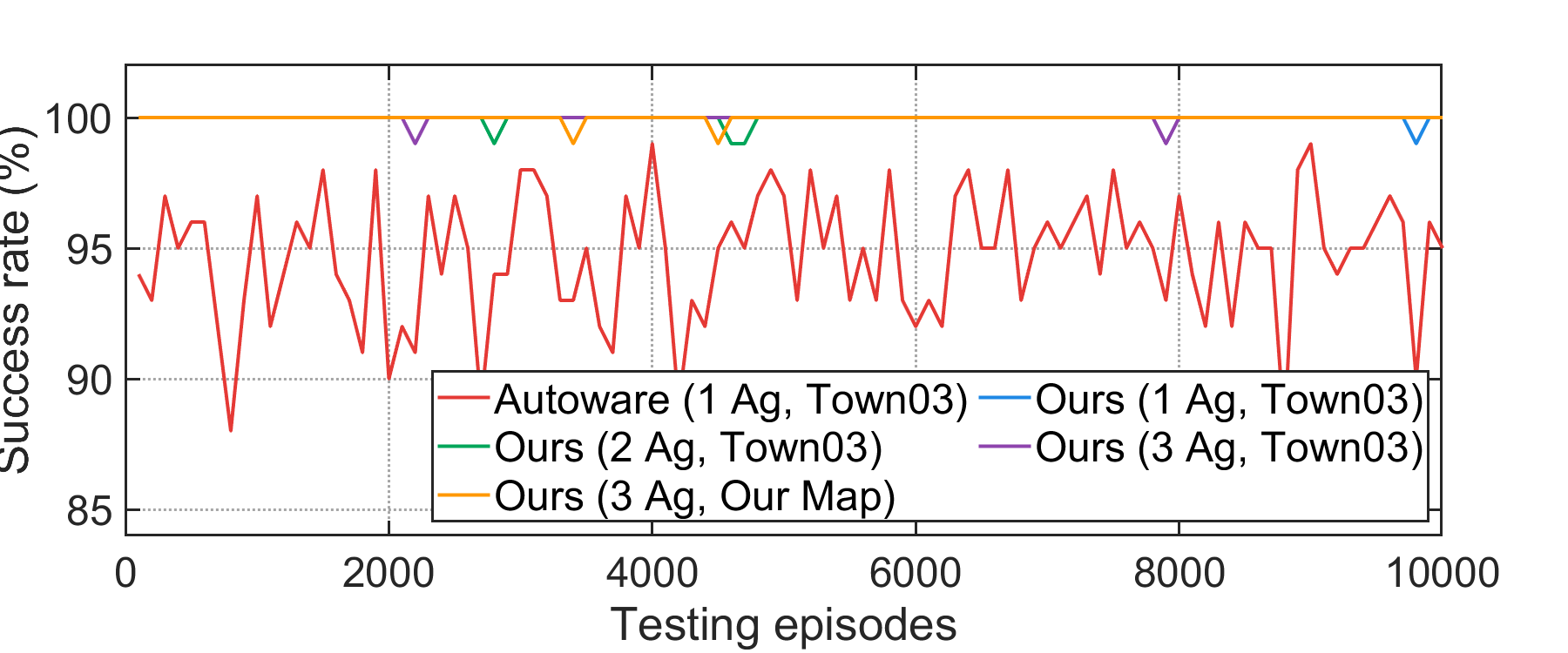} 
\end{minipage}
}
\subfigure[]{
\begin{minipage}[b]{0.45\textwidth}
\includegraphics[width=1\textwidth]{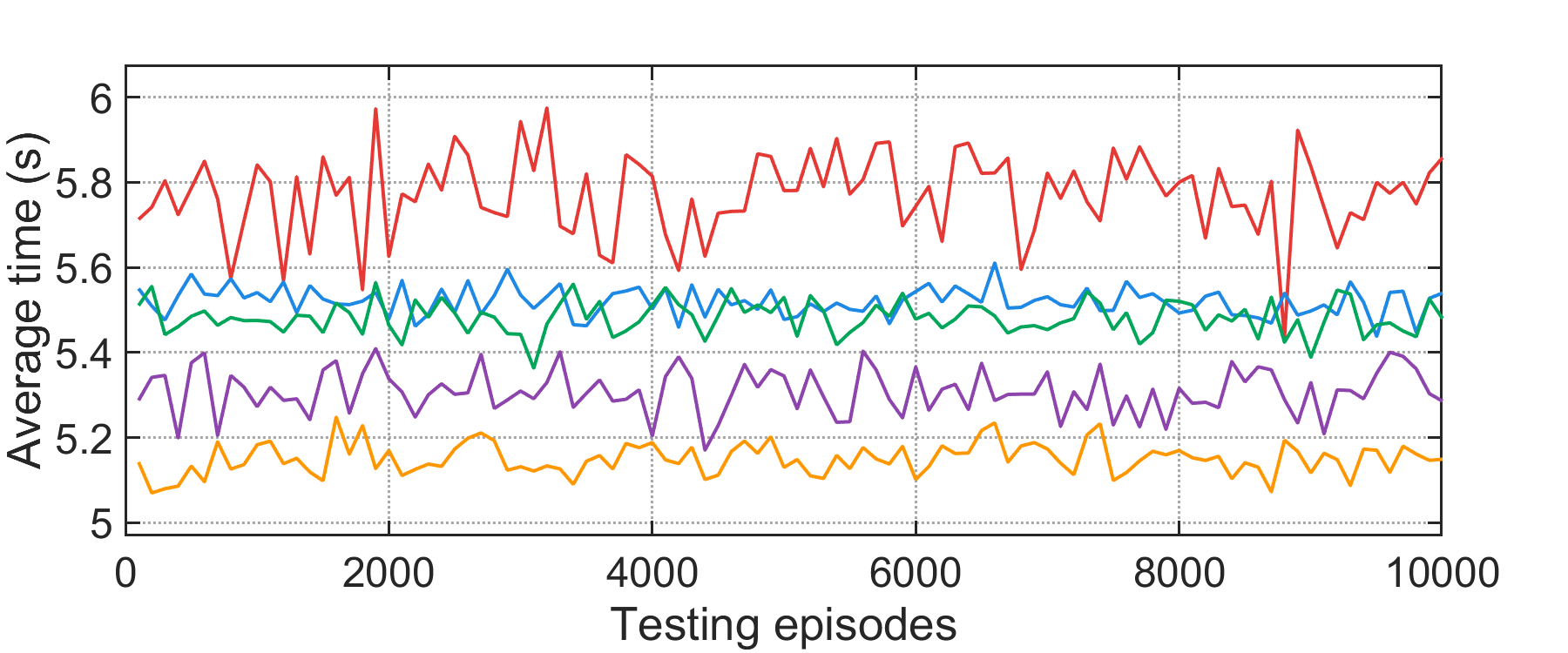} 
\end{minipage}
}
\caption{Final model performance evaluation results (a) success rate by testing episodes, (b) average travel time by testing episodes}
\label{fig:6}
\end{figure}

\subsection{Online Training Results}

Fig.~\ref{fig:5} presents the training convergence of our proposed model alongside two ablated variants. The full model consistently outperforms all baselines. It reaches stable performance within approximately 250 episodes, whereas the online-only baseline requires over 800 episodes to converge. The ablated variant without self-attention and role-specific policies, despite benefiting from offline pre-training, converges after about 500 episodes. This comparison reveals that both components are essential for achieving optimal performance. Offline pre-training accelerates learning and improves initial performance, while self-attention and role-specific policies further enhance and sustain multi-agent coordination effectiveness. These results confirm that our hybrid method combines the safety of offline RL with the adaptability needed for complex multi-agent coordination at intersections.

\subsection{Performance Evaluation and Generalization Analysis}

We evaluated the model through 10,000 performance test episodes on both Town03 intersection and the real intersection map, comparing it against baselines. Key performance indicators are summarized in Fig.~\ref{fig:6} and Table.~\ref{tab:performance_comparison_simple}, where failure rates represent the percentage of episodes ending in collision or timeout. Our model demonstrates high safety and reliability across all test scenarios on Town03. When controlling single vehicle, it achieves 0.01\% failure rate, outperforming the 5.31\% failure rate of the Autoware baseline. Notably, as coordination complexity increases, our system does not exhibit a marked decline in performance. Specifically, the failure rate is 0.03\% in the two-vehicle scenario and 0.02\% in the three-vehicle scenario. The combination of the BEV perspective and the self-attention mechanism contributes to this robustness, demonstrating our model's effectiveness in handling complex multi-agent cooperative tasks.

In addition, the performance advantage of our model is also demonstrated in terms of traffic efficiency. The average travel time in the single-vehicle scenario was 5.52 seconds, compared to the 5.77 seconds by the Autoware. As the number of controlled vehicles increased, the average travel time slightly decreases, indicating that multi-agents coordinated effectively, establishing efficient cooperative driving strategies that actually improve intersection throughput with more CAVs.

For generalization, the three-vehicle model trained on Town03 is deployed on the real intersection map. It achieves a failure rate of 0.02\% and an average travel time of 5.15 seconds in this new environment. This suggests that the BEV effectively eliminates the impact of individual vehicle blind spots. This result validates the excellent generalization capability of our model and provides a solid foundation for the practical application of the approach.

To further validate the system’s deployability, we evaluate its computational performance across different coordination scenarios. Experiments are conducted on an NVIDIA RTX 4070 Ti GPU at a 10 Hz control frequency. The average inference times are 23.7 ms for single vehicle, 31.4 ms for two vehicles, 38.2 ms for three vehicles, and a maximum inference time of 42.6 ms across all tests. This sub-linear scaling confirms efficient multi-agent processing. Even in worst cases, inference time stays well within the 100 ms control interval, leaving sufficient margin for V2I communications and safety checks, validating the system’s real-time deployability.

\section{Conclusion and Future Works}

In this paper, we have proposed a DT-based cooperative driving system with RSU-centric architecture at unsignalized intersections. The system leverages BEV perception to eliminate blind spots and employs a hybrid reinforcement learning algorithm for robust multi-agent cooperative driving strategies. We developed role-specific policies and validated the system in diverse scenarios, achieving a 0.03\% failure rate and sub-40ms inference time for up to three CAVs. Future primary work involves the proof-of-concept (PoC) experiments to fully validate the system performance in the real world.


\begin{thebibliography}{00}
\bibitem{b1} K. Chu, A. Lam and V. Li, "Traffic Signal Control Using End-to-End Off-Policy Deep Reinforcement Learning," in \emph{IEEE Transactions on Intelligent Transportation Systems}, vol. 23, no. 7, pp. 7184-7195, 2022.

\bibitem{b2} U.S. Department of Transportation. [Online]. Available: https://highways .dot.gov /sites/fhwa.dot.gov/files/2024-08

\bibitem{b3} National Highway Traffic Safety Administration. [Online]. Available: https://w ww.nhtsa.gov/press-releases/nhtsa-2023-traffic-fatalities-estimate-april-2024

\bibitem{b4} K. Wang et al., "Smart Mobility Digital Twin Based Automated Vehicle Navigation System: A Proof of Concept," in \emph{IEEE Transactions on Intelligent Vehicles}, vol. 9, no. 3, pp. 4348-4361, 2024.

\bibitem{b5} O. Hashash, C. Chaccour, W. Saad, T. Yu, K. Sakaguchi and M. Debbah, "The Seven Worlds and Experiences of the Wireless Metaverse: Challenges and Opportunities," in \emph{IEEE Communications Magazine}, vol. 63, no. 2, pp. 120-127, 2025.

\bibitem{b6} K. Wang, C. She, Z. Li, T. Yu, Y. Li, and K. Sakaguchi, "Roadside Units Assisted Localized Automated Vehicle Maneuvering: An Offline Reinforcement Learning Approach," in \emph{2024 IEEE 27th International Conference on Intelligent Transportation Systems (ITSC)}, pp. 1709--1715, 2024.

\bibitem{b7} Z. Li, K. Wang, T. Yu, and K. Sakaguchi, "Het-SDVN: SDN-Based Radio Resource Management of Heterogeneous V2X for Cooperative Perception," \emph{IEEE Access}, vol. 11, pp. 76255-76268, 2023.

\bibitem{b8} D. Suo, B. Mo, J. zhao, and S. E. Sarma, "Proof of Travel for Trust-Based Data Validation in V2I Communication," \emph{IEEE Internet of Things Journal}, vol. 10, no. 11, pp. 9565--9584, 2023.

\bibitem{b9} Y. Cui, H. Xu, J. Wu, Y. Sun and J. Zhao, "Automatic Vehicle Tracking With Roadside LiDAR Data for the Connected-Vehicles System," \emph{IEEE Intelligent Systems}, vol. 34, no. 3, pp. 44--51, 2019.

\bibitem{b10} L. Wang et al., "Multi-Modal 3D Object Detection in Autonomous Driving: A Survey and Taxonomy," in \emph{IEEE Transactions on Intelligent Vehicles}, vol. 8, no. 7, pp. 3781-3798, 2023.

\bibitem{b11} K. Moller, R. Trauth, and J. Betz, "Overcoming Blind Spots: Occlusion Considerations for Improved Autonomous Driving Safety," in \emph{2024 IEEE Intelligent Vehicles Symposium (IV)}, pp. 819-826, 2024.

\bibitem{b12} Y. Zhu, Z. He and G. Li, "A bi-Hierarchical Game-Theoretic Approach for Network-Wide Traffic Signal Control Using Trip-Based Data," \emph{IEEE Transactions on Intelligent Transportation Systems}, vol. 23, no. 9, pp. 15408-15419, 2022.

\bibitem{b13} M. Gallo, "Combined Optimisation of Traffic Light Control Parameters and Autonomous Vehicle Routes," \emph{Smart Cities}, no. 3, pp. 1060--1088, 2024.

\bibitem{b14} Y. Shi, H. Dong, C. He, Y. Chen and Z. Song, "Mixed Vehicle Platoon Forming: A Multiagent Reinforcement Learning Approach," \emph{IEEE Internet of Things Journal}, vol. 12, no. 11, pp. 16886-16898, 2025.

\bibitem{b15} S. Iqbal and F. Sha, "Actor-Attention-Critic for Multi-Agent Reinforcement Learning," \emph{arXiv preprint arXiv:1810.02912}, 2019.

\bibitem{b16} R. Younas, H. M. Raza Ur Rehman, I. Lee, B. W. On, S. Yi and G. S. Choi, "SA-MARL: Novel Self-Attention-Based Multi-Agent Reinforcement Learning With Stochastic Gradient Descent," in \emph{IEEE Access}, vol. 13, pp. 35674-35687, 2025.

\bibitem{b17} K. Wang, T. Yu, Z. Li, K. Sakaguchi, O. Hashash, and W. Saad, "Digital Twins for Autonomous Driving: A Comprehensive Implementation and Demonstration," in \emph{2024 International Conference on Information Net-working (ICOIN)}, pp. 452--457, 2024.

\bibitem{b18} A. Dosovitskiy, G. Ros, F. Codevilla, A. Lopez, and V. Koltun, "CARLA: An Open Urban Driving Simulator," in \emph{Proceedings of the 1st Annual Conference on Robot Learning}, pp. 1--16, 2017.

\bibitem{b19} J. Bock, R. Krajewski, T. Moers, S. Runde, L. Vater and L. Eckstein, "The inD Dataset: A Drone Dataset of Naturalistic Road User Trajectories at German Intersections," in \emph{2020 IEEE Intelligent Vehicles Symposium(IV)}, p. 1929--1934,2020.

\bibitem{b20} A. Kumar, A. Zhou, G. Tucker, and S. Levine, "Conservative Q-Learning for Offline Reinforcement Learning," \emph{arXiv preprint arXiv:2006.04779}, 2020.

\bibitem{b21} D. A. Pomerleau, "ALVINN: An Autonomous Land Vehicle in a Neural Network," in \emph{Advances in Neural Information Processing Systems}, 1988.

\bibitem{b22} C. Yu, A. Velu, E. Vinitsky, J. Gao, Y. Wang, A. Bayen, and Y. Wu, "The Surprising Effectiveness of PPO in Cooperative, Multi-Agent Games," \emph{arXiv preprint arXiv:2103.01955}, 2022.

\bibitem{b23} Autoware. [Online]. Available: https://autoware.org/
\end{thebibliography}
\end{document}